\documentclass[superscriptaddress,amsmath,amssymb,aps,prl,twocolumn,reprint,floatfix]{revtex4}
\usepackage{graphicx}
\usepackage{amssymb,amsmath,bbm,braket,indentfirst,bm}
\usepackage{dcolumn}
\usepackage{color,soul}
\usepackage{bigstrut}

\def \g{\gamma}    \def \a{\alpha}
\def \w{\omega}     
\def \s{\sigma}      
\def \e{\epsilon}   \def \r{\rho} 
   \def \d{\delta} 
    \def \l{\lambda} 
      
 \def \t {\theta}

\def \G{\Gamma}    
\def \Lam{\Lambda} 

\def \>{\rangle} 
\def \<{\langle} 
 
\newcommand{\be}{\begin{equation}}
\newcommand{\ee}{\end{equation}}
\newcommand{\beq}{\begin{eqnarray}}
\newcommand{\eeq}{\end{eqnarray}}

\begin{document}

\title{Thermodynamics of statistical inference by cells}

\author{Alex~H.~Lang}
\affiliation{Physics Department, Boston University, Boston, Massachusetts 02215, USA}

\author{Charles~K.~Fisher}
\affiliation{Physics Department, Boston University, Boston, Massachusetts 02215, USA}

\author{Thierry Mora}
\affiliation{Laboratoire de physique statistique, CNRS,
UPMC and \'Ecole normale sup\'erieure,  75005 Paris, France}

\author{Pankaj~Mehta}
\affiliation{Physics Department, Boston University, Boston, Massachusetts 02215, USA}


\begin{abstract}
The deep connection between thermodynamics, computation, and information is now well established both theoretically and experimentally. Here, we extend these ideas to show that thermodynamics also places fundamental constraints on statistical estimation and learning. To do so, we investigate the constraints placed by (nonequilibrium) thermodynamics on the ability of biochemical signaling networks to estimate the concentration of an external signal. We show that accuracy is limited by energy consumption, suggesting that there are fundamental thermodynamic constraints on statistical inference.
\end{abstract}

\maketitle

Cells often perform complex computations in response to external signals. These computations are implemented using elaborate biochemical networks that may operate out of equilibrium and consume energy \cite{Berg1977Physics,Bialek2005Physical, Endres2009Maximum, Hu2010Physical, Mora2010Limits, Sourjik2012Responding,Kaizu2014Berg}. Given that energetic costs place important constraints on the design of physical computing devices \cite{Landauer1961Irreversibility} and neural computing architectures \cite{Laughlin2001Energy}, one may conjecture that thermodynamic constraints also influence the design of cellular information processing networks. This raises interesting questions about the relationship between the information processing capabilities of biochemical networks and energy consumption \cite{Mehta2012Energetic,Lan2012The-energy-speed-accuracy,Govern2012Fundamental, Govern2013How-biochemical, Barato2013Information}.  Indeed, we will show that thermodynamics places fundamental constraints on the ability of biochemical networks to perform statistical inference. More generally, statistical inference is intimately tied to the manipulation of information and hence offers a rich setting to study the  relationship between information and thermodynamics \cite{Berut2012Experimental,Mandal2012Work,Vaikuntanathan2011Modeling, Sagawa2012Nonequilibrium, Still2012Thermodynamics}.

In order for a cell to formulate an appropriate response to an environmental signal, it must first estimate the concentration of an external signaling molecule using membrane bound receptors \cite{Berg1977Physics, Bialek2005Physical, Endres2009Maximum, Mora2010Limits, Hu2010Physical, Sourjik2012Responding, Cheng2013Large}. The biophysics and biochemistry of cellular receptors is highly variable. Whereas some simple receptor proteins behave like two-state systems (i.e.\ unbound and ligand bound) with dynamics obeying detailed balance \cite{Keymer2006Chemosensing}, other receptors, such as G-protein coupled receptors (GPCRs), can actively consume energy as they cycle through multiple states. This naturally raises questions about how energy consumption by cellular receptors affects their ability to perform statistical inference. Here, we address these questions by  analyzing the accuracy of statistical inference (i.e.\ learning) as a function of energy consumption in a simple but biophysically realistic model. We show that learning more accurately always requires expending more energy, suggesting that the accuracy of a statistical estimator is fundamentally constrained by thermodynamics. 

\begin{figure}[b]
\begin{center}
  \includegraphics[angle=0,width=.45\textwidth]{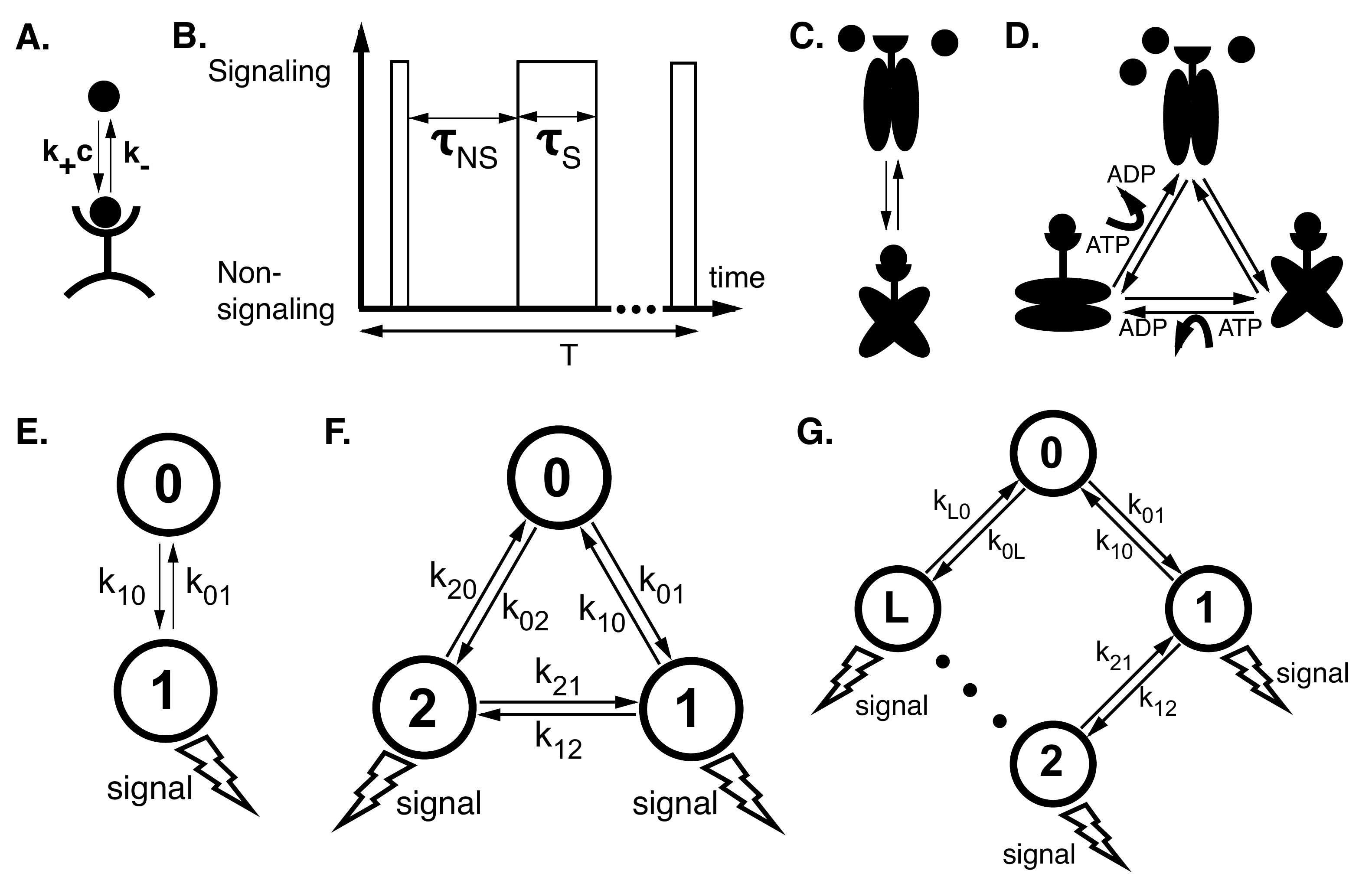}
  \caption{\label{fig:Fig_1} Schematic of a cell receptor and our model of a receptor.  (A) A chemical ligand at concentration $c$ binds to the receptor at rate $k_+ c$ and unbinds at rate $k_-$. (B) Example time series of a receptor binding. While unbound, the receptor is in nonsignaling state, but upon ligand binding it transitions to a signaling state. After a long time $T$, the receptor has a series of nonsignaling times $\tau_{NS}$ and signaling times $\tau_S$ from which to estimate the concentration. (C) Two-state and (D) three-state biochemical models of a receptor. Upon ligand binding the receptor undergoes a physical change (represented as a conformational change) that transmits signals to the downstream biochemical network. (E) Two-state, (F) three-state, and (G) $L$-state Markov models of a receptor, where the chain of states $3$, $4$, $\ldots$ $L-1$ has been suppressed.}
\end{center}
\end{figure}

Cells estimate the concentration of an external ligand using ligand-specific receptors expressed on the cell surface. A ligand (usually a small molecule), at a concentration $c$ in the environment, binds the receptor at a concentration-dependent rate, $k_+c$,  and unbinds at a concentration-independent rate, $k_-$  \cite{Berg1977Physics} (see Fig. 1A). Upon ligand binding, the receptor protein undergoes conformational changes or chemical modifications that alter its activity, sending a signal that the ligand is bound to downstream portions of the biochemical network. During a time interval $T$, the receptor can undergo multiple stochastic transitions between the unbound nonsignaling state and the bound signaling states. This information is contained in the time series of signaling and nonsignaling intervals (see Fig. 1B). After a time $T$, the cell converts this time series into an estimate for the external concentration. A longer time series $T$ always gives a better estimate for the concentration; however the cell needs to make a decision in a finite time, so we consider $T$ to be fixed to a large but finite value. In principle, the estimate for the concentration could be computed using one of many different statistics that can be obtained from this time series (e.g. average bound time, average unbound time, etc.). Each of the resulting estimators for the external ligand concentration has a different accuracy. Following Berg and Purcell (BP) \cite{Berg1977Physics}, we measure the accuracy of an estimator for the concentration using its ``uncertainty,'' defined as:
\be
\mbox{uncertainty} := \frac{\<(\delta c)^2 \>}{\overline{c}^2}
\ee 
where $\overline{c}$ is the mean and $\<(\delta c)^2 \>$ is the variance of the estimated concentration. 

Several methods have been proposed for how a cell may estimate the concentration of the external signaling molecule. In their pioneering paper, Berg and Purcell  suggested  estimating the concentration using the average time the receptor was bound during the time $T$ \cite{Berg1977Physics}. They showed that the minimal uncertainty a receptor could achieve with this estimator was
\be
{\<(\delta c)^2 \> \over \overline{c}^2} =  \frac{2}{\overline{N}}
\label{BP}
\ee
where $\overline{N}$ is the expected  number of binding events during the time interval $T$. For 30 years, many thought that the BP estimator placed a fundamental limit on the accuracy of a cellular receptor. However, in 2009, Endres and Wingreen \cite{Endres2009Maximum} showed that  a cell using maximum likelihood estimation (MLE) based on the average nonsignaling time could reduce its uncertainty by half to
\be
{\<(\delta c)^2 \> \over \overline{c}^2} =  \frac{1}{\overline{N}}.
\label{MLE}
\ee
However, the increased accuracy of MLE comes at an energetic cost. Previous work \cite{Mora2010Limits} established that BP sets a limit for the best possible estimator in equilibrium, implying that any receptor that performs MLE must operate out of equilibrium and consume energy.
	
In order to study the relationship between thermodynamics and the accuracy of statistical estimators, we introduce a new family of biophysically inspired cellular receptors that interpolate between BP and MLE. In our model, receptors can actively consume energy by operating out of equilibrium (for example by hydrolyzing adenosine triphosphate or ATP). Using this family of models, we show that there is a direct connection between the energy  consumed by a receptor and the uncertainty of the resulting estimator. We find that in order to learn more information (decrease its uncertainty), the receptor must always expend more energy (increase entropy production).  Note that, in this paper, we restrict ourselves to modeling the  receptor and ignore the downstream signaling network that converts the signal from the receptor into a cellular response \cite{Mehta2012Energetic,Govern2013How-biochemical}. Thus, the energies computed here represent lower bounds on the total energy consumed by the statistical estimation network.

Fig. 1C shows the simple two-state receptor considered by BP. The binding of an external ligand to the receptor induces a change in the receptor  from a nonsignaling state to a signaling state (see Fig 1B). The dynamics of this simple two-state receptor always obey detailed balance. Thus, in order to model nonequilibrium receptors, we must consider receptors with more than two states. Fig 1D shows a receptor with three states: one nonsignaling state to which ligands can bind, and two signaling states to which ligands cannot bind. With this extra state, the dynamics of the receptor can break detailed balance by coupling the conformational change in the receptor to another reaction such as the hydrolosis of ATP. In particular, by consuming energy it is possible to drive the system preferentially through a series of state changes \cite{Hill1989Free}, for example clockwise in Fig 1F and Fig 1G.  This results in a nonzero probability flux through the state space and positive entropy production.

In order to relate the thermodynamic properties of these receptors to their ability to perform statistical inferences, it is useful to represent receptors as Markov chains. For example, the two-state receptor shown in  Fig. 1C can be represented as a two-state Markov chain with a state $0$ corresponding to the unbound nonsignaling state and state $1$ corresponding to the signaling state (see Fig. 1E). We choose the transition rates between states in the Markov chain to be identical to the transition rates between conformations of the receptor. The three-state receptor can also be modeled as a three-state Markov chain with a ring structure, with state $0$ once again corresponding to the unbound, nonsignaling state (Fig. 1F). In this more abstract notation, it is easy to generalize the three-state receptor considered above to a receptor with $L+1$ states (see Fig. 1G):  $L$ of these states are signaling states that cannot bind the ligand, while the remaining state, $0$, corresponds to the nonsignaling state that can bind ligands. For ease of analysis, in this paper, we consider receptors arranged in a ring only. However, our model is a good approximation for more complicated receptors with multiple pathways, so long as the receptor has a single path (for example, of length $L^*$) that dominates the probability flux, see \cite{Murugan2012Speed} for details. In that case, the complicated receptor reduces to a single ring of length $L^*$.

A straightforward calculation shows that for the architectures in Fig 1 \cite{SI_ref}, the uncertainty of an estimate for the concentration is given by \cite{Endres2009Maximum}:
\be
{\<(\delta c)^2 \> \over \overline{c}^2} =  \frac{1}{\overline{N}}\left[ 1+   \frac{ \<(\delta \tau_S)^2 \> }{ \overline{\tau}_S^2 }  \right] \equiv \frac{\mathcal E}{\overline{N}}
\label{GeneralUncertainty}
\ee
where $N$ is the number of binding events, $\overline{\tau}_S$ is the mean time spent in the signaling state after binding a ligand, and $\<(\delta \tau_S)^2 \>$ is the variance of the time spent in the signaling states. In the second equality, we have defined the coefficient $\mathcal E$ which measures the accuracy of an estimator; e.g.\  ${\mathcal E}=2$ for the Berg-Purcell limit and ${\mathcal E}=1$ for MLE. For a given estimator (i.e. a specific architecture and a set of rates $\vec{k}$), we can calculate the mean and the variance of the signaling time by a first passage calculation similar to that in \cite{Bel2010The-simplicity,SI_ref}. 

Here we provide some intuition for Eq. (\ref{GeneralUncertainty}). Notice that all the information about the ligand concentration is contained in the event of a ligand binding to the receptor, and the unbinding of the ligand, or the exiting of the signaling state, is independent of concentration. Thus, any variation in the duration of the signaling state adds additional noise to the estimate but does not contain any more information about the concentration. Therefore, the optimal estimator is one where the signaling intervals are completely deterministic and  $\<(\delta \tau_S)^2 \>=0$. Comparing Eqs.\ (\ref{GeneralUncertainty}) and (\ref{MLE}), we see that this corresponds to MLE. This is consistent with the well-known fact that MLE is the optimal unbiased estimator for large sample sizes. When the durations of the signaling times are exponentially distributed, like for a two-state receptor, $\<(\delta \tau_S)^2 \>=\overline{\tau}_S^2$, then Eq. (\ref{GeneralUncertainty}) reduces to the BP result given in Eq. (\ref{BP}). Finally, in all cases, the  uncertainty  scales inversely with the average number of binding events $\overline{N}$ during the time interval $T$. This scaling law follows from the central limit theorem by treating each binding event as an independent sample of the concentration.

The Markov representation allows us to calculate the energy consumption using ideas from nonequilibrium statistical physics. We focus on  long time intervals,  $T \gg 1$, with many binding events, where the receptor dynamics can be modeled by nonequilibrium steady states (NESS). The entropy production of the Markov process is the energy per unit time (power) required to maintain this NESS, and therefore calculating the entropy production is equivalent to calculating the energy consumed by the biochemical network\cite{Hill1989Free,Mehta2012Energetic}. The entropy production is given by \cite{Lebowitz1999A-Gallavotti}
\be
e_p = \sum_{i=0}^L \sum_{j \neq i}^L p_i^{ss} k_{ij} \ln \frac{k_{ij}}{k_{ji}},
\ee
with $p_i^{ss}$ is the steady state probability of state $i$,  $k_{ij}$ is the transition rate from state $i$ to state $j$, and we have set $k_BT=1$ \cite{SI_ref}. For the architectures where the Markov process forms a ring,  the entropy production simplifies to
\be
e_p = \left( p_0^{ss} k_{01} - p_1^{ss} k_{10} \right)  \ln \frac{k_{01}k_{12} ... k_{L0}}{k_{0L}k_{10}...k_{L,L-1}} = J \ln \gamma
\ee
where $J$ is the net flux around the ring and $\ln \gamma$ is the free energy per cycle \cite{Hill1989Free,SI_ref}. For later reference, the total energy released in ATP hydrolysis is approximately 20 $k_B T$ at room temperature\cite{Voet2004Biochemistry}. We note that previous work investigating trade-offs between accuracy and energy in Markov chains used a non-thermodynamically feasible energy \cite{Escola2009Maximally}.

Our goal is to find the best performing estimator for a given receptor architecture and entropy production (energy consumption) rate. However, there are several biological constraints that need to be considered when optimizing over choices of kinetic parameters. First, the rate at which a chemical ligand binds to a receptor is set by diffusion limited binding \cite{Berg1977Physics} and hence $k_{01}$ is not controlled by the cell. Therefore we set $k_{01}=1$ and do not optimize over this rate. Second, a receptor needs to be specific. In principle, both ``correct'' ligands (i.e. the ligands the receptor has evolved to detect) and ``wrong'' ligands (any other chemical) can bind the receptor. However, nonspecific ligands quickly unbind and cause the receptor to switch back to the nonsignaling state. Thus, the specificity of a receptor is set by the mean duration of the signaling state in the presence of the correct ligand, $\overline{\tau}_S$. This is incorporated by requiring a small nonspecific binding rate ($k_{0L}=\e \ll 1  = k_{01}$) and we do not optimize over $k_{0L}$. Lastly, since any statistical estimator is always improved with more samples, to fairly compare different families of estimates, we will fix the sampling rate, $\overline{n}=\overline{N}/T$, where $\overline{N}$ is the expected number of samples and $T$ is the signal integration time. By fixing the nonspecific binding rate ($k_{0L}$) to be small \cite{SI_ref}, this implies $\overline{\tau}_S \approx \overline{n}^{-1} -1$. But since we are also fixing the sampling rate, $\overline{n}$, this fixes  $\overline{\tau}_S$. In summary, our goal is to find the global minima for uncertainty, given the above constraints. 

\begin{figure}[b]
\begin{center}
  \includegraphics[angle=0,width=0.45\textwidth]{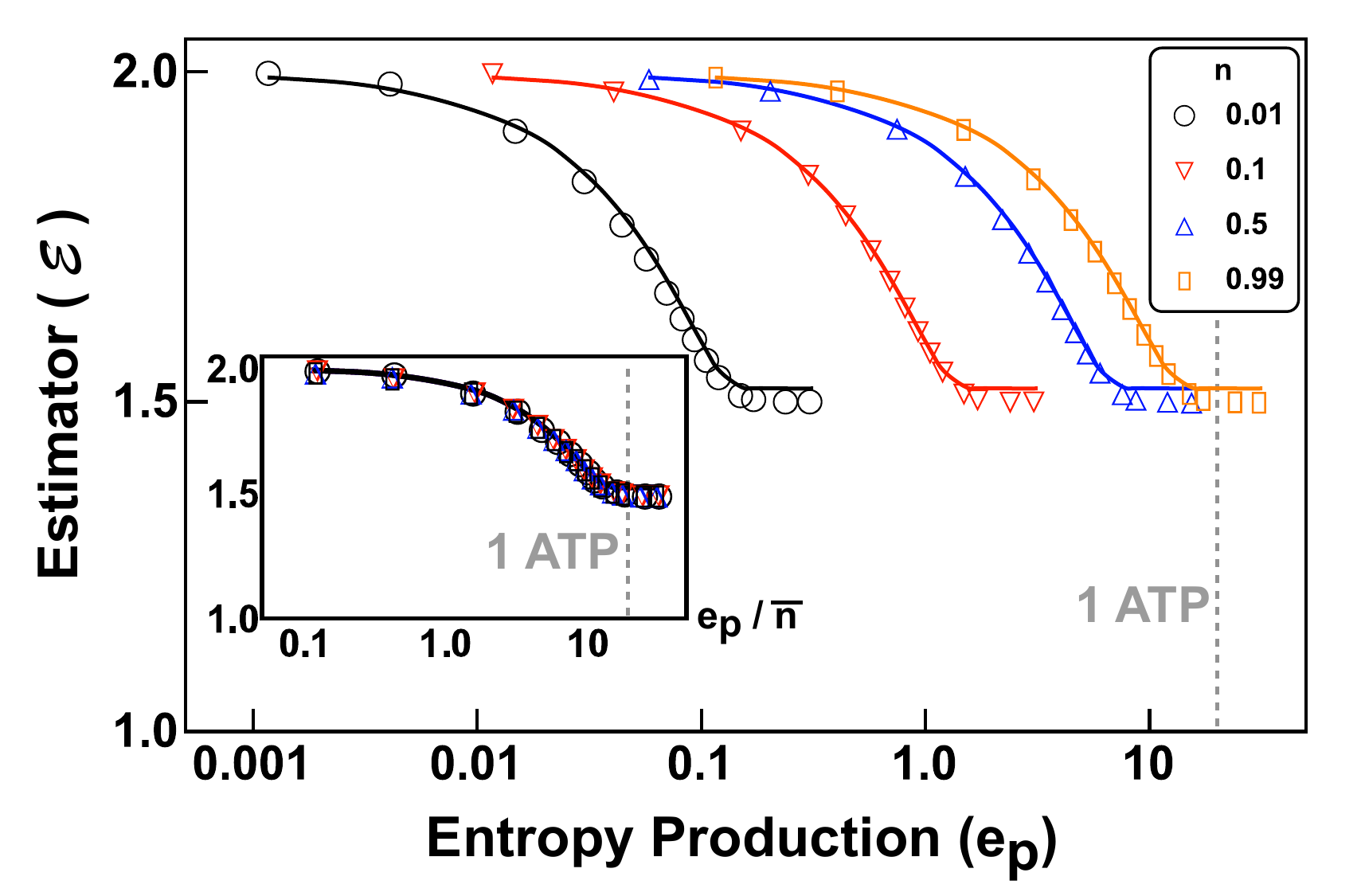}
  \caption{\label{fig:Fig_3}   Two signaling state estimator performance. For varying sampling rate $\overline{n}=\overline{N}/T$, the plot shows estimator performance ($\mathcal{E}$) versus entropy production ($e_p$ with units of $k_BT=1$). The symbols represent results from simulated annealing, where $k_{01}=1$ and $k_{02}=\epsilon=10^{-3}$ while the other four rates are optimized. The continuous lines represent our ansatz\cite{SI_ref} for the global minima. At high entropy production the estimators asymptotically approach $1.5$. The inset shows the data collapse when the estimator performance ($\mathcal{E}$) is plotted versus entropy production per sampling rate ($e_p/\overline{n}$). The vertical dashed line corresponds to the approximate energy released by hydrolysis of a single ATP.}
\end{center}
\end{figure}

We begin by analyzing the three-state receptor (Fig. 1F). Figure 2 shows the uncertainty as a function of entropy production for the optimal threeistate receptor for four different choices of the  ligand binding rate, $\bar{n}=\overline{N}/T$.  To generate these plots, we have used an analytic ansatz\cite{SI_ref} for the optimal parameters  which we have checked using simulated annealing (with agreement within $1.25\%$).  Notice that  learning more accurately (reducing uncertainty) always increased energy consumption (entropy production). At low energy consumption, the receptor approaches the equilibrium BP limit ($\mathcal{E}=2$), while at high energy consumption (corresponding approximately to the energy of ATP hydrolysis) the optimal performance asymptotically approaches the infinite entropy production analytic limit of 
\be
\frac{\<(\delta c)^2 \> }{ \overline{c}^2} \sim \frac{3}{2\overline{N}}
\ee

One striking observation is that these curves exhibit a data collapse when plotted as a function of the energy consumption per ligand binding rate, $e_p/\overline{n}$ . The inset of Fig. 2 shows the same curves as the main graph as a function of $e_p/\overline{n}$. Since each ligand binding event can be viewed as an independent sample of the external concentration, this data collapse suggests that the natural variable linking thermodynamics and inference is the energy per independent sample consumed in constructing an estimator. 
\begin{figure}[t]
\begin{center}
  \includegraphics[angle=0,width=0.45\textwidth]{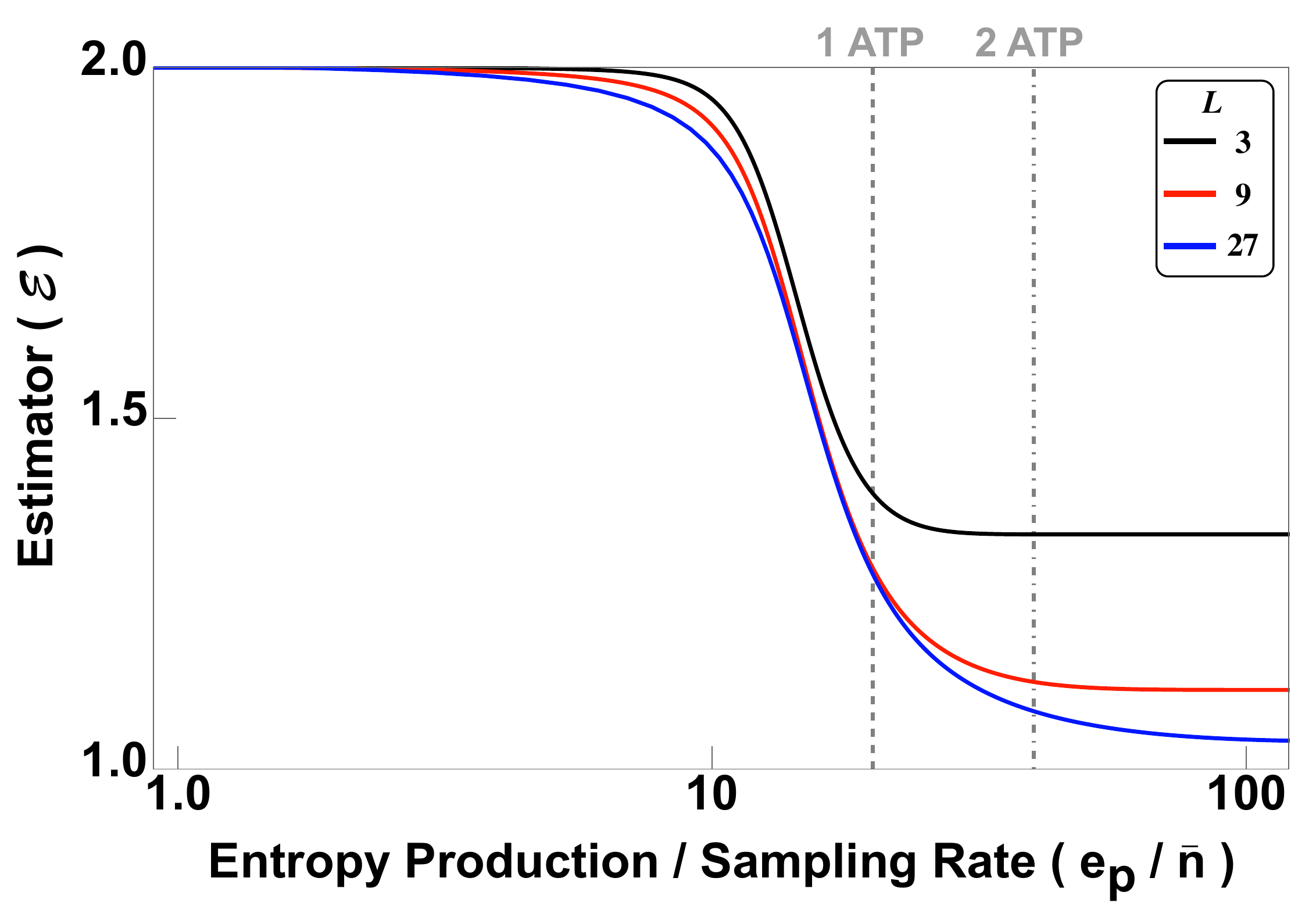}
  \caption{\label{fig:Fig_4} Illustrative example of $L$ signaling state estimator performance. For a varying number of signaling states $L$, the plot shows estimator performance ($\mathcal{E}$) versus energy consumption ($e_p/\overline{n}$). For an increasing $L$, at high energy consumption the estimator approaches the maximum likelihood limit of $1$. The following parameters are fixed at $\overline{n}=0.99$, $k_{01}=1$, $k_{0L} = 10^{-3}$, $\a=k_{10}/b=10^{-3}$, and $\w=k_{L0}/f=1$, while $b$ was varied to keep $\overline{n}$ fixed, and $\t=f/b$ was varied to change the estimator and the energy consumption. These parameters were chosen for convenience and are not global optima. The vertical lines correspond to the approximate energy released by hydrolysis of a single ATP (dashes) or two ATPs (dot dashes).}
\end{center}
\end{figure}

The three-state receptor is not able to reach the MLE limit of $\mathcal{E}=1$ for any level of entropy production. To reach the MLE limit, we consider a receptor with $L+1$ states, $L$ of which are signaling states (see Fig.\ 1G).  This Markov chain has $2L$ independent parameters, which makes it hard to find the global optimum. For this reason, we analyzed a simplified, but still biophysically realistic, rate structure (without performing any optimization over parameters) where $k_{01}$, $k_{0L} $, $k_{10}$, $k_{L0}$ can independently vary but all other forward rates are fixed to be identical,  $k_{i,i+1}=f$   and all other backward rates chosen so that  $k_{i+1,i}=b$, where $i=1 \ldots L-1$ \cite{SI_ref}. Once again, for all choices of $L$,  the optimal uncertainty exhibits a data collapse as a function of the energy consumption per ligand binding rate, $e_p/\overline{n}$ (see Fig. 3). At low energy consumption, the uncertainty approaches the BP limit ($\mathcal{E}=2$), while at high energy consumption (corresponding approximately to the energy of ATP hydrolysis) asymptotically approaches the infinite entropy production analytic limit of 
 \be
\frac{\<(\delta c)^2 \> }{ \overline{c}^2} \sim \left(1+\frac{1}{L} \right)\frac{1}{\overline{N}}
\ee
Thus, receptors with large energy consumption and many signaling states ($L\gg1$) approach the MLE limit. In order to perfectly achieve the MLE limit, all backward rates $b$ would need to be $0$, leading to infinite entropy production. An interesting feature of these curves is that beyond some scale (which can be achieved by hydrolysis of only a few ATP), the marginal gain in improvement that results from consuming more energy becomes negligible. This is reminiscent of the recently found transition in kinetic proofreading where adding additional energy only marginally improves the error threshold \cite{Munsky2009Specificity,Murugan2012Speed}. It will be interesting to see if this is a generic feature of many biochemical information processing circuits.

In conclusion, by analyzing the ability of cells to estimate the concentration of an external chemical signal using nonequilibrium receptors we have established an unexpected link between statistical inference and thermodynamics. Specifically, we found that the efficacy of an estimator for the concentration of a ligand depends on the energy consumed per independent sample by the receptor. Extrapolating this result suggests that there may be fundamental thermodynamic bounds on statistical inference.  The trade-off between accuracy and energy is general and may be relevant for other signal transduction systems, such as gene regulation \cite{Suter2011Mammalian}, light-activated proteins \cite{Bialek2012Biophysics-Searching} or ligand-gated ion channels \cite{Csanady2010Strict}. We note that following the tradition of Berg and Purcell, in this paper we only considered estimating a concentration after a long time $T$. However, in many related cases, such as transcription \cite{Depken2013Intermittent}, the speed is an important trade-off in addition to accuracy and energy consumption. In the context of phosphorelays, it is likely that the circuits can respond quickly even for multistep cascades. For example, the four-stage phosphorelay utilized for phototransduction in the retina can still respond to stimuli in about half a second \cite{Detwiler2000Engineering}. Nonetheless, understanding these trade-offs represents an important future research direction. 

We conjecture that our observed scaling, ($e_p/\overline{n}$), reflects a general principle: the efficiency of a statistical estimator is limited by the energy consumed per sample during its construction. Of course, much more investigation is needed to see if this conjecture holds in general. In particular, it will be interesting to see if these results change for receptors modeled by heterogeneous Markov networks that are not strictly ringlike in nature. Recent work indicates that at large entropy production the dynamics of such networks may be independent of details of the underlying topology, suggesting that our basic picture should hold even for more complicated nonequilibrium receptors \cite{Vaikuntanathan2013Dynamic}.  An additional extension to our model would be to consider externally varying concentrations by implementing a sensory adaptive system (SAS) as was done in recent papers\cite{Sartori2014Thermodynamic,Barato2014Efficiency}. These papers found that the accuracy and energy consumption of the SAS depends on the time scale of external concentration fluctuations. Finally, it is well known that many receptors, such as GPCRs, actively consume energy in order to operate. Our model presents one possible explanation for this observation. The energy consumption may help reduce noise in the downstream signal, allowing cells to more accurately determine external concentrations. Our model also shows that hydrolysis of only one or two ATP nearly achieves the theoretical minima of uncertainty. This may explain why cell sensors often require only a few phosphorylation sites.

We would like to thank  David Schwab and Javad Noorbakhsh for the useful discussions. We also thank Luca D'Alessio and DJ Strouse for detailed comments on the paper. AHL was supported by a National Science Foundation Graduate Research Fellowship (NSF GRFP) under Grant No. DGE-1247312. PM and CKF were supported by a Sloan Research Fellowship and a NIH K25 GM086909.

\newpage

\onecolumngrid
\appendix

\part{Supplementary Material}

\section{Notation}
Here we provide more details of the results in the main text. First, we outline our notation. The time dependent probability of state $i$ is $p_i=p_i(t)$, while the steady state probability of state $i$ is $p_i^{ss}$.  The Laplace transformed probability of state $i$ is $P_i(s)$. The rate to go from state $i$ to state $j$ is $k_{ij}$. The probability to transition from state $i$ to state $j$ is $q_{ij}$. The time it takes to transition from state $i$ to $j$ is $\tau_{ij}$. The first passage time is given by $f(t)$ while the Laplace transformed first passage time is $F(s)$. The lifetime of state $i$ is $\rho_i$.

\section{Detailed Derivation of General Uncertainty}
Here we derive formulas for the accuracy of statistical inference when the activated signaling states continuously produce signals. Following Berg and Purcell \cite{Berg1977Physics}, we will measure the accuracy of a receptor by the ``uncertainty'' of the concentration estimate:
\be
\mbox{uncertainty} := {\<(\delta c)^2 \> \over \overline{c}^2}
\ee 
where $\overline{c}$ is the mean and $\<(\delta c)^2 \>$ is the variance of the estimated concentration. 

Let us consider the case where activated signaling states produce downstream signaling molecules at a rate $\alpha$. We will define $\overline{\tau}_S$ as the mean lifetime of the signaling states and $\overline{\tau}_{NS}$ as the mean non-signaling time. Then, we know that the mean number of signaling molecules $\overline{u}$ produced after a time $T$   is given by
\be
\overline{u}=    \alpha T \left( { \overline{\tau}_S \over \overline{\tau}_S+ \overline{\tau}_{NS}} \right) \equiv   \alpha T \bar{\phi}
\label{meanu}
\ee
This follows by noting that $\bar{\phi}$ is just the fraction of time the receptor is in the signaling states. Notice that by definition, $\a$ and $T$ are independent of the concentration $c$. The signaling time $\overline{\tau}_S$, can in principle depend on concentration, and for $L$ signaling states is given by
\be
\overline{\tau}_S = \sum_{i=1}^L p_{0i} \tau_{i0}
\ee
where $p_{0i}$ is the probability to transition from state $0$ to state $i$, and $\tau_{i0}$ is the mean time to return from state $i$ to state $0$. Since we assume the receiving state is strongly biased (i.e. $k_{01}$ is much larger than any other rate $k_{0i}$ from non-signaling, $0$, to signaling state $i$), then the derivative of the signaling time with respect to concentration is:
\be
\frac{d\overline{\tau}_S}{dc} = -\sum_{i=2}^L \frac{k_{0i}}{k_{01}} \left( \tau_{10} + \tau_{i0} \right)
\ee
Since this is by assumption small, we will approximate $\overline{\tau}_S$  as independent of concentration, and thus all the concentration dependence comes from $\overline{\tau}_{NS}$. Thus, using the usual error-propagation formulas one has
\be
{\delta u \over \overline{u}} =    -  {d \overline{\tau}_{NS} \over dc} {1 \over \overline{\tau}_S + \overline{\tau}_{NS}} \delta c
\ee
which gives the uncertainty for the concentration:
\be
{\<(\delta c)^2 \> \over \overline{c}^2} =  \left(\overline{c} {d \overline{\tau}_{NS} \over dc}\right)^{-2}  \left(\overline{\tau}_{NS} + \overline{\tau}_{S}  \right)^2{ \<(\delta u)^2 \> \over \overline{u}^2} 
\label{deltaceq}
\ee

The formula above reduces the problem to calculating the uncertainty in the number of signaling molecules produced in a time $T$. To calculate this, notice that $\overline{u}$ comes from on average $\overline{N}= T/(\overline{\tau}_S + \overline{\tau}_{NS})$ independent binding cycles (state $0$ to state $1$ transition). Thus, the variance in the fraction of time bound during a time $T$  will just be $\overline{N}^{-1}$ times the variance in a single binding cycle.  In particular, the coefficient of variation in a single cycle is given by 
\be
\frac{\delta \phi}{\overline{\phi}} =     \frac{\overline{\tau}_{NS}}{\overline{\tau}_S + \overline{\tau}_{NS}} \left[ \left({ \delta \tau_S \over \overline{\tau}_S }\right) - \left({ \delta \tau_{NS} \over \overline{\tau}_{NS} } \right) \right] 
\ee
Noting that the signaling and non-signaling events are independent, we get 
\be
\frac{\< (\delta u)^2 \>}{\overline{u}^2} =    \frac{1}{\overline{N}} \left(\frac{\overline{\tau}_{NS}}{\overline{\tau}_S + \overline{\tau}_{NS}} \right)^2  \left[ { \<(\delta \tau_{NS})^2 \> \over \overline{\tau}_{NS}^2 } +{ \<(\delta \tau_S)^2 \> \over \overline{\tau}_S^2 }  \right] 
\ee

Plugging this expressions into (\ref{deltaceq}) gives
\be
{\<(\delta c)^2 \> \over \overline{c}^2} =  \frac{1}{\overline{N}} \left(\overline{c} {d \log{( \overline{\tau}_{NS})} \over dc}\right)^{-2}\left[ { \<(\delta \tau_{NS})^2 \> \over \overline{\tau}_{NS}^2 } +{ \<(\delta \tau_S)^2 \> \over \overline{\tau}_S^2 }  \right] 
\ee

Therefore the complicated response of a receptor is reduced to its mean and variance of the time in both the signaling and non-signaling states. In this paper, we will examine the case where there is a single non-signaling state ($0$) and there are $L$ signaling states arranged in a ring. In this case, the above expression simplifies to (leading order $k_{0L}/k_{01}$):
\be
{\<(\delta c)^2 \> \over \overline{c}^2} =  \frac{1}{\overline{N}}\left[ 1+   { \<(\delta \tau_S)^2 \> \over \overline{\tau}_S^2 }  \right] 
\ee

 For a two state process as considered by Mora and Wingreen \cite{Mora2010Limits}, there is only the receiving state and one signaling state. These are just Poisson processes which each have an uncertainty of $1$ and we recover the Berg and Purcell \cite{Berg1977Physics} limit
\be
{\<(\delta c)^2 \> \over \overline{c}^2} =  \frac{2}{\overline{N}}
\ee

\section{General First Passage Time}
We need to calculate the first passage properties of the Markov chain, specifically the mean and variance of the first passage time. This can be calculated as follows \cite{Redner2001A-Guide,Bel2010The-simplicity}. The master equation that we want to solve is $\frac{dp}{dt} = Kp(t)$. 

First apply the Laplace transform
\be
P_i(s) = \int_0^\infty p_i(t) e^{-st} dt
\ee
which leads to the master equation
\be
\left(s-K \right) P(s) = p(t=0) 
\ee
with $K$ the matrix of transitions for the full system but with the transition rates leaving the absorbing states set to zero.

The first passage time to return to state $0$ is
\beq
f(t) &=& \frac{dp_0(t)}{dt} \\
F(s) &=& s P_0(s)
\eeq

For our purposes, we only need the mean and variance of the first passage time. This is easily obtained by the uncentered moments
\beq
M^{(m)} &=& \int_0^\infty t^m f(t) = (-1)^m \left. \frac{d^m F(s)}{ds^m} \right|_{s=0}
\eeq
where $m=1$ is the mean and $m=2$ is the uncentered second moment.

In general we know that $\tau_x$, the spent in state $x$, is drawn from a mixture where it can switch to states $j=1,2,...$. The variance of mixtures is $X= \sum_i w_i X_i$, where $w_i$ are arbitrary weights and $X_i$ are random variables drawn from distributions with mean $\mu_i$ and variance $\sigma_i$. Combining equations we get:
\be
\text{Var(X)}= \sum_{i} w_i \left[ (\mu_i -\mu)^2 +  \sigma_i^2\right]
\ee
with $\mu = \sum_i w_i \mu_i$.

We can get the time spent in state $x$, $\tau_x$, by using the variance mixture formula combined with $\overline{\tau}_{ix}$ and $\text{Var}( \tau_{i  x})$, respectively the mean and variance first passage time of starting in state $i$ and ending in state $x$. This gives us
\beq
\overline{\tau}_x &=& \sum_i q_{xi} \overline{\tau}_{i x} \\
q_{x i} &=& \frac{k_{xi}}{\sum_j k_{xj} }=  k_{xi} \rho_{x} \\
 \rho_{x} &=& \left(\sum_j k_{xj} \right)^{-1} \\
\text{Var($\tau_x$)} &=& \sum_i q_{x i}  Var( \tau_{i  x}) +  \sum_{i} q_{x  i}  \left(\overline{\tau}_{i x}- \sum_{k} q_{x k} \overline{\tau}_{k  x} \right)^2 
\eeq
where $q_{xi}$ is the probability of transitioning from state $x$ to state $i$, $k_{xi}$ is the rate to go from state $x$ to state $i$, and $\rho_x$ is the lifetime of state $x$.

In this paper, we have one non-signaling state and the other $L$ states are signaling. Therefore, we will let state $0$ be the absorbing state, and it can initially transition to state $1$ and state $L$. The above equations then simplify to
\beq
\overline{\tau_0} &=& q_{01} \overline{\tau}_{1 0}+ q_{0L} \overline{\tau}_{L0} \\
\text{Var($\tau_0$)} &=& q_{0 1}  Var( \tau_{1 0}) + q_{0 L}  Var( \tau_{L 0}) + 2 q_{0 1} q_{0L}   \left(\overline{\tau}_{1 0}-  \overline{\tau}_{L 0} \right)^2 \\
q_{0 L} &=& 1-q_{0 1}
\eeq

\section{First Passage Time: 2 Signaling States}
Here we calculate the mean and variance of the first passage time to return to state $0$ from either state $1$ or $2$. The master equation that we need to solve is $\frac{dp}{dt} = Kp(t)$. The matrix rates are:
\be
K_{ij} = \begin{cases} 
k_{10} & \mbox{for } i=0 \mbox{ and } j=1\\
k_{12} & \mbox{for } i=2 \mbox{ and } j=1\\
k_{20} & \mbox{for } i=0 \mbox{ and }  j=2\\
k_{21} & \mbox{for } i=1 \mbox{ and } j=2\\
-(k_{10}+k_{12}) & \mbox{for } i=1 \mbox{ and } j=1\\
-(k_{20}+k_{21}) & \mbox{for } i=2 \mbox{ and } j=2\\
0 & \mbox{everywhere else}
\end{cases}
\ee

While the initial conditions are set by the rates $k_{01}$ and $k_{02}$, for the purposes of the first passage time calculation, the rates from $0$ to $1$  ($k_{01}$) and from $0$ to $2$ ($k_{02}$) are both set to zero, $k_{01}=k_{02}=0$.

The Laplace transform for the initial condition of starting in state $1$ is:
\be
F(s) = sP_0(s) = k_{10}P_1+k_{20}P_2
\ee

with 
\beq
P_1 &=& \left[ \G_1 - \frac{k_{12}k_{21}}{\G_2} \right]^{-1} \\
P_2 &=& \frac{k_{12}}{\G_2} P_1 \\
\G_i &=& s + \r_i^{-1}
\eeq

We can obtain mean and variance from
\beq
\overline{\tau} &=& - \left .\frac{dF}{ds}\right|_{s=0} \\
\text{Var($\tau$)}&=&  \left .\frac{d^2F}{ds^2}\right|_{s=0} - \overline{\tau}^2
\eeq

The mean and variance of the first passage time from starting in either state $1$ or state $2$ is:
\beq
\overline{\tau}_{10} &=& \rho_1 \frac{1+k_{12}\rho_2}{1-k_{12}k_{21}\rho_1\rho_2} = \frac{k_{12}+k_{20}+k_{21}}{\xi} \\
\overline{\tau}_{20} &=& \rho_2 \frac{1+k_{21}\rho_1}{1-k_{12}k_{21}\rho_1\rho_2} = \frac{k_{10}+k_{12}+k_{21}}{\xi} \\
\text{Var($\tau_{10}$)} &=& \tau_{10}^2 \left[1 + 2 \r_2^2\frac{k_{12}\left(k_{10}-k_{20} \right)}{(1+k_{12}\r_2)^2} \right]=  \tau_{10}^2+2\frac{k_{12}(k_{10}-k_{20})}{\xi^2}  \\
\text{Var($\tau_{20}$)} &=& \tau_{20}^2 \left[1+ 2\r_1^2 \frac{k_{21}\left(k_{20}-k_{10} \right)}{(1+k_{21}\r_1)^2} \right] =  \tau_{20}^2 - 2\frac{k_{21}(k_{10}-k_{20})}{\xi^2}  \\
\xi &=& k_{10}k_{20}+k_{10}k_{21}+k_{12}k_{20}
\eeq
where the second equality holds as long as $\xi \neq 0$.

\section{First Passage Time: L Signaling States}

\subsection{Derivation}

Here we calculate the mean and variance of the first passage time in a $L+1$ state chain. The master equation that we need to solve is $\frac{dp}{dt} = Kp(t)$. The matrix is indexed from $0$ to $L$ and the rates are:
\be
K_{ij} = \begin{cases} 
k_{10} & \mbox{for } i=0 \mbox{ and } j=1\\
k_{L0} & \mbox{for } i=0 \mbox{ and } j=L\\
f & \mbox{for } i=j+1 \mbox{ and }  1<j<L\\
b & \mbox{for } i=j-1 \mbox{ and } 1<j<L\\
-(f+k_{10}) & \mbox{for } i=1 \mbox{ and } j=1\\
-(f+b) & \mbox{for } i=j \mbox{ and } 1<j<L\\
-(k_{L0}+b) & \mbox{for } i=L \mbox{ and } j=L\\
0 & \mbox{everywhere else}
\end{cases}
\ee

While the initial conditions are set by the rates $k_{01}$ and $k_{0L}$, for the purposes of the first passage time calculation, the rates from $0$ to $1$  ($k_{01}$) and from $0$ to $L$ ($k_{0L}$) are both set to zero, $k_{01}=k_{0L}=0$.

For later convenience we define the following ratio of rates:
\beq
\t &=& \frac{f}{b} \\
\a &=& \frac{k_{10}}{b} \\
\w &=& \frac{k_{L0}}{f}
\eeq

\begin{figure}[b]
\begin{center}
  \includegraphics[angle=0,width=.45\textwidth]{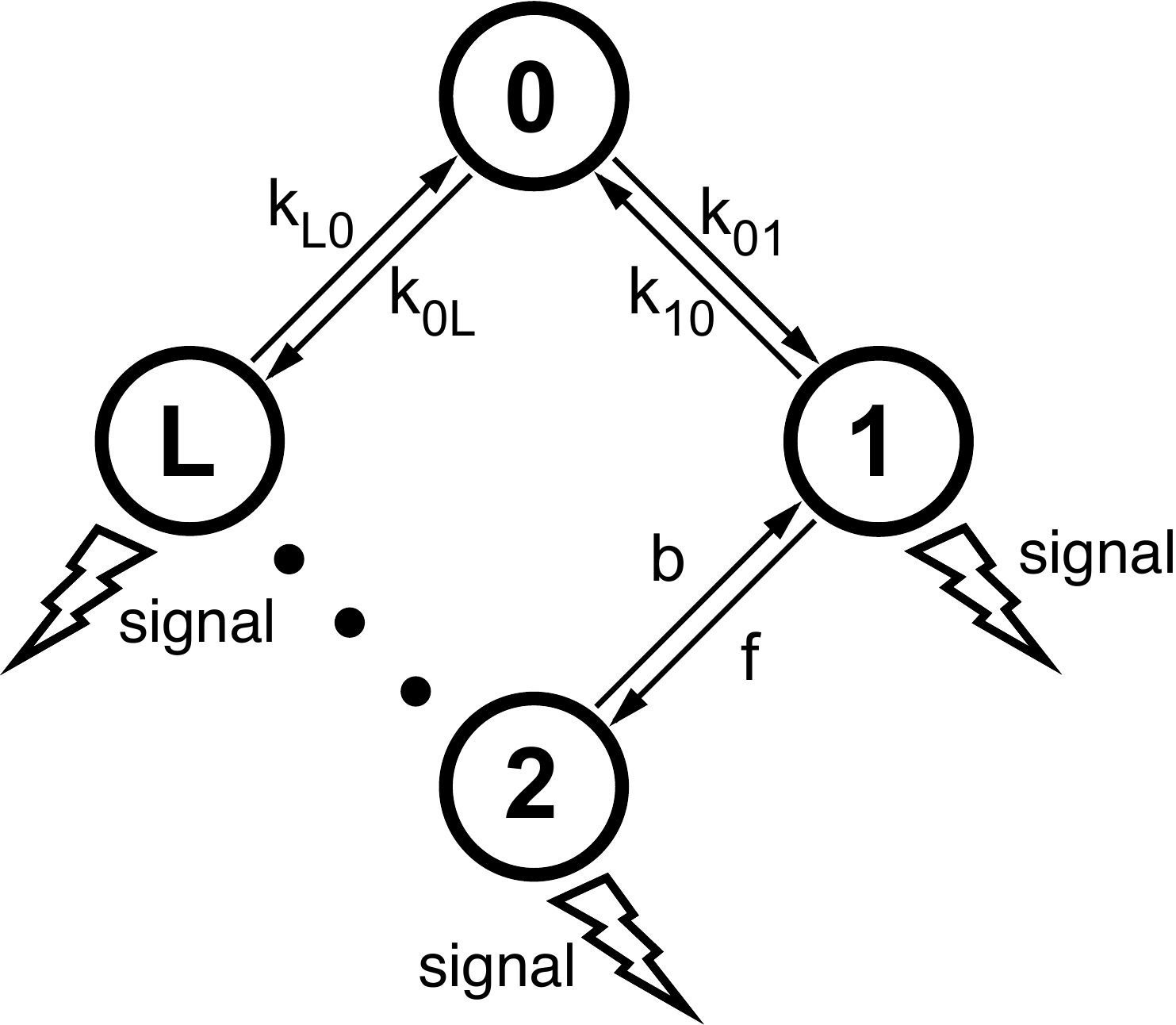}
  \caption{\label{fig:SI_Fig_1} Simplified rate structure considered for $L$ signaling states first passage time calculation. The rates  $k_{01}$, $k_{10}$, $k_{L0}$, $k_{0L}$ are unconstrained, while the remaining forward rates are equal, $f = k_{12} = k_{23} = \ldots =k_{L-1,L} $ and the remaining backward rates are equal, $b = k_{21} = k_{32} = \ldots =k_{L,L-1} $.}
\end{center}
\end{figure}

We can use a transfer matrix to find a general solution (for non-degenerate eigenvales, i.e. $\t \neq 1$) to the state probability as
\beq
P_i(s) &=& C_+ \l_+^{i-1} + C_- \l_-^{i-1}
\eeq

Solving for the the expressions $1<i<L$ leads to 
\beq
\l_\pm &=& \frac{1}{2b}\left( s+f+b \pm \sqrt{(s+f+b)^2-4fb} \right) \\
&=& \frac{1}{2}\left( \sigma \pm \sqrt{\sigma^2-4\t} \right) =  \frac{1}{2}\left( \sigma \pm \psi \right) \\
\s &=& \frac{s}{b} + \t + 1 \\
\psi &=& \sqrt{\sigma^2-4\t}
\eeq

With the initial condition of starting in $P_1$, the boundary equations for $P_1$ and $P_{L}$ are:
\beq
(\s +\a -1 )\left( C_+ + C_- \right) &=& 1/b +  \left( C_+ \l_+ + C_- \l_- \right) \\
(\s +(\w-1)\t)\left( C_+ \l_+^{L-1} + C_- \l_-^{L-1} \right) &=&  \t  \left( C_+  \l_+^{L-2} + C_-  \l_-^{L-2} \right) 
\eeq

Solving these equations gives 
\beq
C_- &=& - C_+ \Lam^{L}M \\
C_+ &=& \frac{1}{b\left[\l_-+\a-1  - (\l_++\a-1)\Lam^{L}M    \right]} \\
\Lam &=& \frac{\l_+}{\l_-} \\
M &=& \frac{1+(\w-1)\l_-}{1+(\w-1)\l_+}
\eeq

And then the probabilities are
\beq
P_1(s) &=& C_+ \left( 1-\Lam^{L}M \right) \\
P_{L}(s) &=& C_+ \l_+^{L-1} \left(1 - \Lam M \right) 
\eeq

The full Laplace transform $F$ is:
\beq
F(s) &=& \frac{\a (1-\Lam^{L}M) + \w \t \l_+^{L-1}(1-\Lam M)}{\l_-+\a-1  - (\l_++\a-1)\Lam^{L}M   }
\eeq

\subsection{Results}

To get the mean and variance of the first passage time, we need
\beq
\overline{\tau}_{10} &=& - \left. \frac{dF}{ds}\right|_{s=0} \\
\text{Var($\tau_{10}$)} &=&  \left. \frac{d^2F}{ds^2}\right|_{s=0} - \overline{\tau}^2 
\eeq

The mean return time to state $0$ when starting in state $1$ is:
\beq
\overline{\tau}_{10} &=& \frac{\overline{\tau}_{10,num}}{\overline{\tau}_{10,den}}\\
\overline{\tau}_{10,num} &=& (\w L -\w +1 )\t^{L+1} -(\w L +1)\t^{L}+(\w-1)\t+1\\
\overline{\tau}_{10,den} &=& b   \left[ \t -1\right] \left[ \w \t^{L+1}+\w(\a-1)\t^{L}+\a(1-\w)\t-\a  \right]
\eeq

The variance of the return time to state $0$ when starting in state $1$ is:
\beq
\text{Var($\tau_{10}$)} &=& \frac{\text{Var($\tau_{10}$)}_{num}}{\text{Var($\tau_{10}$)}_{den}} \\
   \text{Var($\tau_{10}$)}_{num}&=& \t ^{2 L+3} \left[\w ^2 (L-1)+1\right] \\  \nonumber
   &+&\t ^{2 L+2} \left[\w ^2 \left(L^2
   \a -L (3 \a +1)+2 \a -3\right)+2 \w  ((L-2) \a
   +1)+2 \a -3\right]\\ \nonumber
    &-&\t ^{2 L+1} \left[\w ^2
   \left(2 L^2 \a -4 L \a +L+4 \a -4\right)+\w  ((4 L-6)
   \a +4)+4 \a -3\right]\\  \nonumber
   &+&\t ^{2 L} \left[\w  (\w  L+2) (L \a -\a +1)+2 \a -1\right]\\  \nonumber
   &+&
    \t ^{L+3} (\w -1) \left[2 (\w -1)
   \a +3 \w  L^2 \a +L (\w  (4-5 \a )+4 \a
   )+2\right]\\   \nonumber
      &+&\t
   ^{L+2} \left[-2 \w ^2 \left(3 L^2 \a +L (4-6 \a )+\a
   -2\right)\right] \\  \nonumber
   &+&\t
   ^{L+2} \left[\w  \left(9 L^2 \a +L (12-23 \a )+8 \a
   -6\right) + 6(2 L-1) \a +6\right] \\  \nonumber
   &+&\t ^{L+1} \left[\w 
   \left(-9 L^2 \a +L (19 \a -12)-6 \a +6\right)\right]\\  \nonumber
   &+&\t ^{L+1} \left[\w ^2
   (L-1) ((3 L-4) \a +4)+6 (-2 L \a +\a -1)\right] \\  \nonumber
   &+&\t ^L \left[\a  \left(3 L^2 \w -5 L \w +4 L+2
   \w -2\right)+(4 L-2) \w +2\right] \\  \nonumber
   &-& \t ^3 (\w -1)^2  (2 \a -1)\\  \nonumber
   &-& \t ^2 (\w -1) (\w +4
   \a -3)\\  \nonumber
   &+&\t  (-2 \w -2 \a +3)\\  \nonumber
     &-&1\\ 
  \text{Var($\tau_{10}$)}_{den} &=&b^2 \left[ \t -1\right]^3 \left[ \w \t^{L+1}+\w(\a-1)\t^{L}+\a(1-\w)\t-\a  \right]^2 
\eeq

While the results here are for initial condition of being in state $1$, one can easily find the results for the initial condition of state $L$ if one makes the following substitutions $\t \Leftrightarrow 1/\t$, $b \Leftrightarrow f$, and $\a\Leftrightarrow \w$.

\section{Steady State Probabilities}
In general, we are considering a Markov chain with $L+1$ nodes (labeled $0$ to $L$). We have the master equation
\be
\frac{dP(t)}{dt} = KP(t)
\ee
with $K$ the matrix of transition rates. The rates are labeled as $k_{ij}$ where $i$ is the initial state and $j$ is the final state. For later convenience, define the lifetime of a state as 
\be
\r_i = \left(\sum_{j\neq i}k_{ij} \right)^{-1}
\ee

The steady state distributions are easily obtained by solving $Kp^{ss}=0$. The solution can be written in a compact form \cite{Blythe2001Nonequilibrium} as
\beq
P_i^{ss} &=& \frac{z_i}{Z} \\
Z &=& \sum_i z_i
\eeq
and $z_i$ is the matrix minor of $K$ at $(i,i)$ i.e. the determinant of $K$ with the $i$th row and column removed.

For the two signaling state system we have that
\beq
p_0^{ss} &=&\frac{\r_1^{-1}\r_2^{-1}-k_{12}k_{21}}{Z} = \frac{k_{10} k_{20}+ k_{10} k_{21} + k_{12} k_{20}}{Z} \\
p_1^{ss} &=&\frac{\r_0^{-1}\r_2^{-1}-k_{02}k_{20}}{Z} =  \frac{ k_{01} k_{20} + k_{01} k_{21} + k_{02}k_{21}}{Z} \\
p_2^{ss} &=& \frac{\r_0^{-1}\r_1^{-1}-k_{01}k_{10}}{Z} =  \frac{ k_{01}k_{12} + k_{02}k_{10} + k_{02}k_{12} }{Z} \\
Z &=& \sum_{i \neq j} \left(  \r_i^{-1}\r_j^{-1} - k_{ij}k_{ji} \right) 
\eeq

For the $L$ signaling state with the simplified rates, we will just present the result for state $0$:
\beq
p_0^{ss} &=& \frac{p_{0,num}^{ss}}{p_{0,den}^{ss}} \\
p_{0,num}^{ss}&=& b (\t -1) \left(   \w \t^{L+1}+\w(\a-1)\t^{L}+\a(1-\w)\t-\a \right) \\
p_{0,den}^{ss} &=&  -\a  \e +\a  b +\a  L \e +\e +1 \\
   &+&    \t  \left(\a  b
   \w -2 \a  b-\a L \e +\w -\e -1\right) \\
&+&\a b \t ^2 \left(1- \w \right) \\
     &+& \t^L \left( b \w +\a  \e  -L \w -1-\e  -\a  b \w \right) \\
        &+&    \t^{1+L}  \left(\a  b \w  -2 b \w  +L \w  -\w  +1+\e  \right) \\   
&+& b \w  \t ^{L+2} 
\eeq
The rates from $0$ to $1$ is $k_{01}=1$, from $1$ to $0$ is $k_{10}$ (with $\a=k_{10}/b$), from $0$ to $L$ is $k_{0L}=\e \ll 1$, and from $L$ to $0$ is $k_{L0}$ (with $\w=k_{L0}/f$). All other forward rates are $f$ and backward rates are $b$ and the ratio of rates is $\t=f/b$.

\section{Average Sampling Rate: $\overline{n}$}

The average sampling rate is
\be
\overline{n} = \frac{N}{T} = k_{01}p_0^{ss}
\ee
where $N$ is the number of samples (i.e. number of binding events), $T$ is the total integration time, $k_{01}$ is the rate from state $0$ to state $1$, and $p_0^{ss}$ is the steady state probability of being in state $0$.

Since we are assuming that $k_{01}=1$ and $k_{L0}=\e \ll 1$, we have the mean signaling time becomes $\overline{\tau}_S \approx \tau_{10}$. With these rates we have 
\be
\overline{n} \approx \left(1+\overline{\tau}_S \right)^{-1}
\ee

\section{Entropy Production: $e_p$}
For a general Markov process with states labeled by $i$, steady state probabilities $p_i^{ss}$, and transition rate $k_{ij}$ from state $i$ to state $j$, the non-equilibrium steady state (NESS) entropy production \cite{Lebowitz1999A-Gallavotti, Mehta2012Energetic} is given by
\be
e_p = \sum_{i=0}^L  \sum_{j \neq i}^L p_i^{ss} k_{ij} \ln \frac{k_{ij}}{k_{ji}}
\ee
where the summation is over both $i$ and $j$. Alternatively, the entropy production can be written as a sum over the flux between each connected node as 
\be
e_p =\sum_{i=0}^L \sum_{j>i}^L \left( p_i^{ss} k_{ij} - p_i^{ss} k_{ij} \right)  \ln \frac{k_{ij}}{k_{ji}}
\ee
where now we have an unrestricted sum over $i$ but a restricted sum over $j$.

Since we are modeling our receptor as a ring,  the entropy production simplifies to
\be
e_p = \left( p_0^{ss} k_{01} - p_1^{ss} k_{10} \right)  \ln \frac{k_{01}k_{12} \ldots k_{L0}}{k_{0L}k_{10} \ldots k_{L,L-1}} = J \ln \gamma
\ee
where the flux $J = p_0^{ss} k_{01} - p_1^{ss} k_{10}$ between each neighboring state is equal and the $\ln \gamma$ is the free energy difference of a cycle.

For $2$ signaling states, the entropy production per sampling rate is given by:
\beq
\frac{e_p}{\overline{n}} &=& \left[1+\frac{k_{10}}{k_{12}} + \frac{k_{10} k_{21}}{k_{12} k_{20}}\right]^{-1} \frac{\g-1}{\g} \ln \g \\
\g &=& \frac{k_{01}k_{12}k_{20}}{k_{10}k_{21}k_{02}}
\eeq

For the $L$ signaling states arranged in a ring, the entropy production per sampling rate is given by:
\beq
\frac{e_p}{\overline{n}} &=& \left[1+\frac{\a}{\w}\t^{-L} +\a \t^{-1} \frac{1-\t^{1-L}}{1-\t^{-1}} \right]^{-1} \frac{\g-1}{\g} \ln \g \\
\g &=& \frac{k_{01}\w }{k_{0L }\a}\t^L
\eeq
where $\w=k_{L0}/f$, $\a=k_{10}/b$, $\t=f/b$, $f$ is all the forward rates (except $k_{01}$ and $k_{L0}$), and $b$ is all the backward rates (except $k_{10}$ and $k_{0L}$).

\section{Ansatz for 2 Signaling State Receptor}

\begin{figure}[b]
\begin{center}
  \includegraphics[angle=0,width=.45\textwidth]{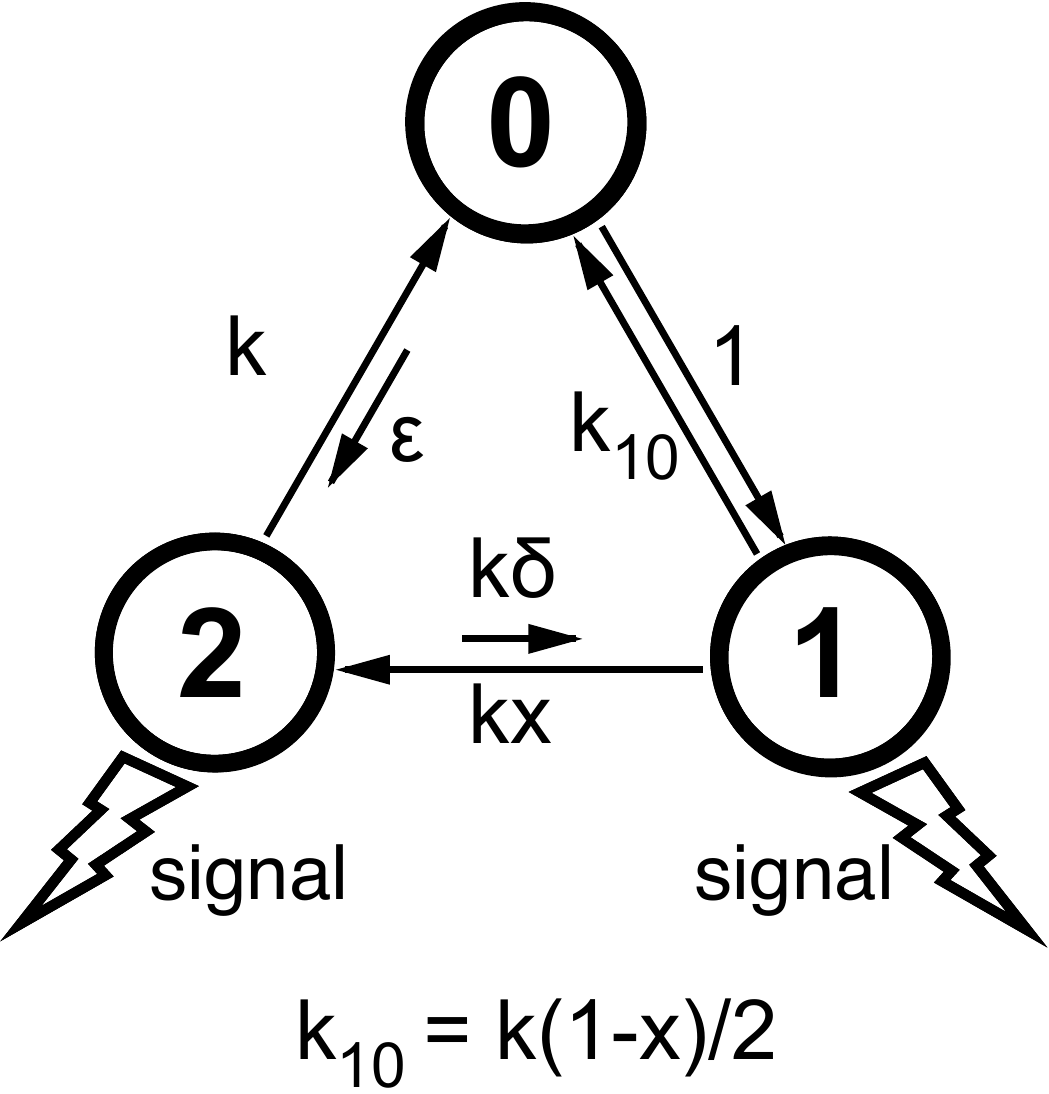}
  \caption{\label{fig:SI_Fig_2} Rate structure for ansatz of minimum uncertainty for the $L=2$ signaling state system. The rates are as follows: $k_{01}=1$, $k_{10}=\frac{k}{2}(1-x)$, $k_{12}=kx$, $k_{21}=k \delta $, $k_{20}=k$,  and $k_{02}=\epsilon$. The mean signaling time is set by $k$. The other rates are $\e, \d \ll 1$ and $0 < x < 1 $.}
\end{center}
\end{figure}

Here are the details of the ansatz for the minimum uncertainty for the 2 signaling state system.

The rates are as follows:
\begin{itemize}
\item $k_{01} = 1$
\item $k_{10} = \frac{k}{2}\left(1-x \right)$
\item $k_{12} = kx$
\item $k_{21} = k\delta$
\item $k_{20} = k$
\item $k_{02} = \epsilon$
\end{itemize}
where $\epsilon \ll 1$ (and in this paper $\epsilon =10^{-3}$), $0 < x < 1$, $\delta \ll 1$ (and in this paper $\delta =0.04$), and $k$ is varied to fix the mean sampling rate $\overline{n}$.

For the ansatz, the mean, coefficient of variation, and entropy production simplifies to
\beq
 \overline{\tau}_S  &\approx& \frac{2}{k} \\
{ \<(\delta \tau_S)^2 \> \over \overline{\tau}_S^2 }  &\approx& 1 - \frac{x}{1+x} \\
\frac{e_p}{\overline{n}} &\approx& \left( 1 + \frac{1-x}{2x} \right)^{-1} \frac{\gamma-1}{\gamma} \ln \g \\
\g &=& \frac{2x}{\e \d(1-x)}
\eeq

\section{Simulated Annealing}

Simulated annealing is a meta-heuristic algorithm for global optimization in which one uses the Metropolis algorithm to perform a random walk in parameter space while periodically lowering the temperature. We used a simulated annealing algorithm to search for the parameters of a model describing a receptor with 2 signaling states that minimizes a cost function given by
\beq
\text{cost} = \frac{ \< (\d c)^2\>} { \bar{c}^2 }+ \lambda_{e_p} (\ln e_p - \ln \hat{e}_p)^2 - ( \lambda_n\hat{n} - 1) \ln \overline{n} - (\lambda_n(1-\hat{n})-1) \ln (1- \overline{n})
\eeq
That is, we minimize the uncertainty of the resulting estimator ($ \< (\d c)^2 \> / \bar{c}^2$) subject to soft constraints on the energy production ($e_p$) and sampling rate ($\overline{n}$), which are constrained to $\hat{e}_p$ and $\hat{n}$, respectively. Here, $\lambda_{e_p}$ and $\lambda_n$ implement the constraints. We chose $\lambda_{e_p} = 20$ and $\lambda_n = 20 / \max \{\hat{n}, 1- \hat{n} \}$. 

Let $\Omega_1$ denote a set of parameters describing a receptor with 2 signaling states (i.e.\ all of the various rate constants). A new set of trial parameters $\Omega_2$ was generated in the following way: for each $k \in \Omega_1$ set the corresponding $k' \in \Omega_2$ to $\ln k' = \ln k + \eta$ where $\eta$ is a random variable with from a Normal distribution centered at zero. The width of the Normal distribution was chosen adaptively so that approximately 25\% of the steps were accepted. Making the random perturbations to the logarithm of the rate constants ensures that they are always positive. The trial move was accepted according to the Metropolis criterion with probability $\min[1, \exp( (\text{cost}(\Omega_1) - \text{cost}(\Omega_2))/T )]$. The temperature $T$ was initialized to $T = 10$ and adjusted by $T \leftarrow 0.95 T$ every 2000 steps. The best solution obtained during the chain was stored in $\Omega_B$, and the chain was re-initialized from $\Omega_1 = \Omega_B$ every 2000 steps to prevent the chain from getting stuck in a poor local minimum. This simulated annealing algorithm was run until convergence of $ \< (\d c)^2 \> / \bar{c}^2$, $e_p$ and $n$.

\section{Scaling with Temperature}
In the main text, we worked in the units of $k_BT=1$. However, here we examine the general temperature dependence. Experimentally, it is known that rates of biochemical reactions doubles for every $10\,^{\circ}\mathrm{C}$ \cite{Hussain2014Engineered,Segel1975Enzyme}. Therefore, a general rate $k$ at a temperature $T$ (measured in degrees Celsius) is related to initial rate $k_0$ and initial temperature $T_0$ by:
\be
k = k_0 2^{\frac{T-T_0}{10}}
\ee
 
Now we need to determine the general scaling of various entities in this paper, which is summarized below in terms of a general rate $k$:
\begin{itemize}
\item Mean signaling time, $\overline{\tau}_S \sim k^{-1}$
\item Variance in signaling time, $\<(\delta \tau_S)^2 \>  \sim k^{-2}$
\item Coefficient of variation of signaling time, $\frac{\<(\delta \tau_S)^2 \>}{\overline{\tau}_S^2} \sim 1$
\item Sampling rate,  $\overline{n} \sim k$
\item Uncertainity, $\frac{\<(\delta c)^2 \>}{\overline{c}^2} \sim k^{-1}$
\item Entropy production, $e_p \sim k$
\end{itemize}

While increasing temperature increases both the mean and variance of the signaling time, since the estimator ($\mathcal{E}=1+\frac{\<(\delta \tau_S)^2 \>}{\overline{\tau}_S^2}$) only depends on the coefficient of variation of signaling time, the estimator is independent of temperature. The sampling rate $\overline{n}$ does increase with increasing temperature, and therefore increasing temperature decreases the uncertainty. However, this decrease in uncertainty costs energy. While the free energy per cycle ($\ln \gamma$) remains constant, the probability flux ($J$) is proportional to a rate, and since the entropy production is given by $e_p = J \ln \gamma$, we see that that decrease in uncertainty is directly related to the increase in entropy production.


\begin{thebibliography}{31}
\expandafter\ifx\csname natexlab\endcsname\relax\def\natexlab#1{#1}\fi
\expandafter\ifx\csname bibnamefont\endcsname\relax
  \def\bibnamefont#1{#1}\fi
\expandafter\ifx\csname bibfnamefont\endcsname\relax
  \def\bibfnamefont#1{#1}\fi
\expandafter\ifx\csname citenamefont\endcsname\relax
  \def\citenamefont#1{#1}\fi
\expandafter\ifx\csname url\endcsname\relax
  \def\url#1{\texttt{#1}}\fi
\expandafter\ifx\csname urlprefix\endcsname\relax\def\urlprefix{URL }\fi
\providecommand{\bibinfo}[2]{#2}
\providecommand{\eprint}[2][]{\url{#2}}

\bibitem[{\citenamefont{Berg and Purcell}(1977)}]{Berg1977Physics}
\bibinfo{author}{\bibfnamefont{H.}~\bibnamefont{Berg}} \bibnamefont{and}
  \bibinfo{author}{\bibfnamefont{E.}~\bibnamefont{Purcell}},
  \bibinfo{journal}{Biophys. J.} \textbf{\bibinfo{volume}{20}},
  \bibinfo{pages}{193} (\bibinfo{year}{1977}).
  
  \bibitem[{\citenamefont{Bialek and Setayeshgar}(2005)}]{Bialek2005Physical}
\bibinfo{author}{\bibfnamefont{W.}~\bibnamefont{Bialek}} \bibnamefont{and}
  \bibinfo{author}{\bibfnamefont{S.}~\bibnamefont{Setayeshgar}},
  \bibinfo{journal}{Proc. Natl. Acad. Sci. U.S.A.} \textbf{\bibinfo{volume}{102}},
  \bibinfo{pages}{10040} (\bibinfo{year}{2005}).

\bibitem[{\citenamefont{Endres and Wingreen}(2009)}]{Endres2009Maximum}
\bibinfo{author}{\bibfnamefont{R.}~\bibnamefont{Endres}} \bibnamefont{and}
  \bibinfo{author}{\bibfnamefont{N.}~\bibnamefont{Wingreen}},
  \bibinfo{journal}{Phys. Rev. Lett.} \textbf{\bibinfo{volume}{103}},
  \bibinfo{pages}{158101} (\bibinfo{year}{2009}).

\bibitem[{\citenamefont{Hu et~al.}(2010)\citenamefont{Hu, Chen, Rappel, and
  Levine}}]{Hu2010Physical}
\bibinfo{author}{\bibfnamefont{B.}~\bibnamefont{Hu}},
  \bibinfo{author}{\bibfnamefont{W.}~\bibnamefont{Chen}},
  \bibinfo{author}{\bibfnamefont{W.}~\bibnamefont{Rappel}}, \bibnamefont{and}
  \bibinfo{author}{\bibfnamefont{H.}~\bibnamefont{Levine}},
  \bibinfo{journal}{Phys. Rev. Lett.} \textbf{\bibinfo{volume}{105}},
  \bibinfo{pages}{48104} (\bibinfo{year}{2010}).

\bibitem[{\citenamefont{Mora and Wingreen}(2010)}]{Mora2010Limits}
\bibinfo{author}{\bibfnamefont{T.}~\bibnamefont{Mora}} \bibnamefont{and}
  \bibinfo{author}{\bibfnamefont{N.}~\bibnamefont{Wingreen}},
  \bibinfo{journal}{Phys. Rev. Lett.} \textbf{\bibinfo{volume}{104}},
  \bibinfo{pages}{248101} (\bibinfo{year}{2010}).

\bibitem[{\citenamefont{Sourjik and Wingreen}(2012)}]{Sourjik2012Responding}
\bibinfo{author}{\bibfnamefont{V.}~\bibnamefont{Sourjik}} \bibnamefont{and}
  \bibinfo{author}{\bibfnamefont{N.~S.} \bibnamefont{Wingreen}},
  \bibinfo{journal}{Curr. Opin. Cell Bio.}
  \textbf{\bibinfo{volume}{24}}, \bibinfo{pages}{262} (\bibinfo{year}{2012}).

\bibitem[{\citenamefont{Kaizu et~al.}(2014)\citenamefont{Kaizu, de~Ronde,
  Paijmans, Takahashi, Tostevin, and ten Wolde}}]{Kaizu2014Berg}
\bibinfo{author}{\bibfnamefont{K.}~\bibnamefont{Kaizu}},
  \bibinfo{author}{\bibfnamefont{W.}~\bibnamefont{de~Ronde}},
  \bibinfo{author}{\bibfnamefont{J.}~\bibnamefont{Paijmans}},
  \bibinfo{author}{\bibfnamefont{K.}~\bibnamefont{Takahashi}},
  \bibinfo{author}{\bibfnamefont{F.}~\bibnamefont{Tostevin}}, \bibnamefont{and}
  \bibinfo{author}{\bibfnamefont{P.~R.} \bibnamefont{ten Wolde}},
  \bibinfo{journal}{Biophys. J.} \textbf{\bibinfo{volume}{106}},
  \bibinfo{pages}{976} (\bibinfo{year}{2014}).
  
  \bibitem[{\citenamefont{Landauer}(1961)}]{Landauer1961Irreversibility}
\bibinfo{author}{\bibfnamefont{R.}~\bibnamefont{Landauer}},
  \bibinfo{journal}{IBM J. Res. Dev.}
  \textbf{\bibinfo{volume}{5}}, \bibinfo{pages}{183} (\bibinfo{year}{1961}).

\bibitem[{\citenamefont{Laughlin}(2001)}]{Laughlin2001Energy}
\bibinfo{author}{\bibfnamefont{S.}~\bibnamefont{Laughlin}},
  \bibinfo{journal}{Curr. Opin. Neurobiol.}
  \textbf{\bibinfo{volume}{11}}, \bibinfo{pages}{475} (\bibinfo{year}{2001}).

\bibitem[{\citenamefont{Mehta and Schwab}(2012)}]{Mehta2012Energetic}
\bibinfo{author}{\bibfnamefont{P.}~\bibnamefont{Mehta}} \bibnamefont{and}
  \bibinfo{author}{\bibfnamefont{D.~J.} \bibnamefont{Schwab}},
  \bibinfo{journal}{Proc. Natl. Acad. Sci. U.S.A.}
  \textbf{\bibinfo{volume}{109}}, \bibinfo{pages}{17978}
  (\bibinfo{year}{2012}).

\bibitem[{\citenamefont{Lan et~al.}(2012)\citenamefont{Lan, Sartori, Neumann,
  Sourjik, and Tu}}]{Lan2012The-energy-speed-accuracy}
\bibinfo{author}{\bibfnamefont{G.}~\bibnamefont{Lan}},
  \bibinfo{author}{\bibfnamefont{P.}~\bibnamefont{Sartori}},
  \bibinfo{author}{\bibfnamefont{S.}~\bibnamefont{Neumann}},
  \bibinfo{author}{\bibfnamefont{V.}~\bibnamefont{Sourjik}}, \bibnamefont{and}
  \bibinfo{author}{\bibfnamefont{Y.}~\bibnamefont{Tu}},
  \bibinfo{journal}{Nat. Phys.} \textbf{\bibinfo{volume}{8}},
  \bibinfo{pages}{422} (\bibinfo{year}{2012}).
  
  \bibitem[{\citenamefont{Govern and ten Wolde}(2012)}]{Govern2012Fundamental}
\bibinfo{author}{\bibfnamefont{C.~C.} \bibnamefont{Govern}} \bibnamefont{and}
  \bibinfo{author}{\bibfnamefont{P.~R.} \bibnamefont{ten Wolde}},
  \bibinfo{journal}{Phys. Rev. Lett.} \textbf{\bibinfo{volume}{109}},
  \bibinfo{pages}{218103} (\bibinfo{year}{2012}).
  
  \bibitem[{\citenamefont{Govern and Wolde}(2013)}]{Govern2013How-biochemical}
\bibinfo{author}{\bibfnamefont{C.~C.} \bibnamefont{Govern}} \bibnamefont{and}
  \bibinfo{author}{\bibfnamefont{P.~R.} \bibnamefont{ten Wolde}},
  \bibinfo{journal}{arXiv:1308.1449}  (\bibinfo{year}{2013}).
  
  \bibitem[{\citenamefont{Barato et~al.}(2013)\citenamefont{Barato, Hartich, and
  Seifert}}]{Barato2013Information}
\bibinfo{author}{\bibfnamefont{A.}~\bibnamefont{Barato}},
  \bibinfo{author}{\bibfnamefont{D.}~\bibnamefont{Hartich}}, \bibnamefont{and}
  \bibinfo{author}{\bibfnamefont{U.}~\bibnamefont{Seifert}},
  \bibinfo{journal}{Phys. Rev. E} \textbf{\bibinfo{volume}{87}},
  \bibinfo{pages}{042104} (\bibinfo{year}{2013}).

  \bibitem[{\citenamefont{Berut et~al.}()\citenamefont{Berut, Arakelyan, Petrosyan,
  Ciliberto, Dillenschneider, and Lutz}}]{Berut2012Experimental}
\bibinfo{author}{\bibfnamefont{A.}~\bibnamefont{Berut}},
  \bibinfo{author}{\bibfnamefont{A.}~\bibnamefont{Arakelyan}},
  \bibinfo{author}{\bibfnamefont{A.}~\bibnamefont{Petrosyan}},
  \bibinfo{author}{\bibfnamefont{S.}~\bibnamefont{Ciliberto}},
  \bibinfo{author}{\bibfnamefont{R.}~\bibnamefont{Dillenschneider}},
  \bibnamefont{and} \bibinfo{author}{\bibfnamefont{E.}~\bibnamefont{Lutz}},
  \bibinfo{journal}{Nature}  \textbf{\bibinfo{volume}{483}},
  \bibinfo{pages}{187} (\bibinfo{year}{2012}).

\bibitem[{\citenamefont{Mandal and Jarzynski}(2012)}]{Mandal2012Work}
\bibinfo{author}{\bibfnamefont{D.}~\bibnamefont{Mandal}} \bibnamefont{and}
  \bibinfo{author}{\bibfnamefont{C.}~\bibnamefont{Jarzynski}},
  \bibinfo{journal}{Proc. Natl. Acad. Sci. U.S.A.}
  \textbf{\bibinfo{volume}{109}}, \bibinfo{pages}{11641}
  (\bibinfo{year}{2012}).

\bibitem[{\citenamefont{Vaikuntanathan and
  Jarzynski}(2011)}]{Vaikuntanathan2011Modeling}
\bibinfo{author}{\bibfnamefont{S.}~\bibnamefont{Vaikuntanathan}}
  \bibnamefont{and}
  \bibinfo{author}{\bibfnamefont{C.}~\bibnamefont{Jarzynski}},
  \bibinfo{journal}{Phys. Rev. E} \textbf{\bibinfo{volume}{83}},
  \bibinfo{pages}{061120} (\bibinfo{year}{2011}).

\bibitem[{\citenamefont{Sagawa and Ueda}(2012)}]{Sagawa2012Nonequilibrium}
\bibinfo{author}{\bibfnamefont{T.}~\bibnamefont{Sagawa}} \bibnamefont{and}
  \bibinfo{author}{\bibfnamefont{M.}~\bibnamefont{Ueda}},
  \bibinfo{journal}{Phys. Rev. E} \textbf{\bibinfo{volume}{85}},
  \bibinfo{pages}{021104} (\bibinfo{year}{2012}).

\bibitem[{\citenamefont{Still et~al.}(2012)\citenamefont{Still, Sivak, Bell,
  and Crooks}}]{Still2012Thermodynamics}
\bibinfo{author}{\bibfnamefont{S.}~\bibnamefont{Still}},
  \bibinfo{author}{\bibfnamefont{D.~A.} \bibnamefont{Sivak}},
  \bibinfo{author}{\bibfnamefont{A.~J.} \bibnamefont{Bell}}, \bibnamefont{and}
  \bibinfo{author}{\bibfnamefont{G.~E.} \bibnamefont{Crooks}},
  \bibinfo{journal}{Phys. Rev. Lett.} \textbf{\bibinfo{volume}{109}},
  \bibinfo{pages}{120604} (\bibinfo{year}{2012}).

\bibitem[{\citenamefont{Cheng et~al.}(2013)\citenamefont{Cheng, Merchan,
  Tchernookov, and Nemenman}}]{Cheng2013Large}
\bibinfo{author}{\bibfnamefont{X.}~\bibnamefont{Cheng}},
  \bibinfo{author}{\bibfnamefont{L.}~\bibnamefont{Merchan}},
  \bibinfo{author}{\bibfnamefont{M.}~\bibnamefont{Tchernookov}},
  \bibnamefont{and} \bibinfo{author}{\bibfnamefont{I.}~\bibnamefont{Nemenman}},
  \bibinfo{journal}{Phys. Bio.} \textbf{\bibinfo{volume}{10}},
  \bibinfo{pages}{035008} (\bibinfo{year}{2013}).

\bibitem[{\citenamefont{Keymer et~al.}(2006)\citenamefont{Keymer, Endres,
  Skoge, Meir, and Wingreen}}]{Keymer2006Chemosensing}
\bibinfo{author}{\bibfnamefont{J.~E.} \bibnamefont{Keymer}},
  \bibinfo{author}{\bibfnamefont{R.~G.} \bibnamefont{Endres}},
  \bibinfo{author}{\bibfnamefont{M.}~\bibnamefont{Skoge}},
  \bibinfo{author}{\bibfnamefont{Y.}~\bibnamefont{Meir}}, \bibnamefont{and}
  \bibinfo{author}{\bibfnamefont{N.~S.} \bibnamefont{Wingreen}},
  \bibinfo{journal}{Proc. Natl. Acad. Sci. U.S.A.} \textbf{\bibinfo{volume}{103}},
  \bibinfo{pages}{1786} (\bibinfo{year}{2006}).

\bibitem[]{Hill1989Free}
\bibinfo{author}{\bibfnamefont{T.L.}~\bibnamefont{Hill}},
\bibinfo{title}{\bibfnamefont{\emph{Free energy transduction and biochemical cycle kinetics}}},
(\bibinfo{year}{Springer, 1989}).

\bibitem[{\citenamefont{Murugan et~al.}(2012)\citenamefont{Murugan, Huse, and
  Leibler}}]{Murugan2012Speed}
\bibinfo{author}{\bibfnamefont{A.}~\bibnamefont{Murugan}},
  \bibinfo{author}{\bibfnamefont{D.~A.} \bibnamefont{Huse}}, \bibnamefont{and}
  \bibinfo{author}{\bibfnamefont{S.}~\bibnamefont{Leibler}},
  \bibinfo{journal}{Proc. Natl. Acad. Sci. U.S.A.}
  \textbf{\bibinfo{volume}{109}}, \bibinfo{pages}{12034}
  (\bibinfo{year}{2012}).

\bibitem[{\citenamefont{SI}(2014)\citenamefont{SI}}]{SI_ref}
  \bibinfo{journal}{See Supplementary Material for additional details}.

\bibitem[{\citenamefont{Bel et~al.}(2010)\citenamefont{Bel, Munsky, and
  Nemenman}}]{Bel2010The-simplicity}
\bibinfo{author}{\bibfnamefont{G.}~\bibnamefont{Bel}},
  \bibinfo{author}{\bibfnamefont{B.}~\bibnamefont{Munsky}}, \bibnamefont{and}
  \bibinfo{author}{\bibfnamefont{I.}~\bibnamefont{Nemenman}},
  \bibinfo{journal}{Phys. Bio.} \textbf{\bibinfo{volume}{7}},
  \bibinfo{pages}{016003} (\bibinfo{year}{2010}).
  

\bibitem[{\citenamefont{Lebowitz and Spohn}(1999)}]{Lebowitz1999A-Gallavotti}
\bibinfo{author}{\bibfnamefont{J.}~\bibnamefont{Lebowitz}} \bibnamefont{and}
  \bibinfo{author}{\bibfnamefont{H.}~\bibnamefont{Spohn}},
  \bibinfo{journal}{J. Stat. Phys.}
  \textbf{\bibinfo{volume}{95}}, \bibinfo{pages}{333} (\bibinfo{year}{1999}).


\bibitem[{\citenamefont{Voet and Voet}(2004)}]{Voet2004Biochemistry}
\bibinfo{author}{\bibfnamefont{D.}~\bibnamefont{Voet}} \bibnamefont{and}
  \bibinfo{author}{\bibfnamefont{J.}~\bibnamefont{Voet}},
  \emph{\bibinfo{title}{Biochemistry}} (\bibinfo{publisher}{John Wiley and Sons
  New York}, \bibinfo{year}{2004}), \bibinfo{edition}{3rd ed.} \bibinfo{page}{(pg 566, Table 16.3)}


\bibitem[{\citenamefont{Escola et~al.}(2009)\citenamefont{Escola, Eisele,
  Miller, and Paninski}}]{Escola2009Maximally}
\bibinfo{author}{\bibfnamefont{S.}~\bibnamefont{Escola}},
  \bibinfo{author}{\bibfnamefont{M.}~\bibnamefont{Eisele}},
  \bibinfo{author}{\bibfnamefont{K.}~\bibnamefont{Miller}}, \bibnamefont{and}
  \bibinfo{author}{\bibfnamefont{L.}~\bibnamefont{Paninski}},
  \bibinfo{journal}{Neural computation} \textbf{\bibinfo{volume}{21}},
  \bibinfo{pages}{1863} (\bibinfo{year}{2009}).

\bibitem[{\citenamefont{Munsky et~al.}(2009)\citenamefont{Munsky, Nemenman, and
  Bel}}]{Munsky2009Specificity}
\bibinfo{author}{\bibfnamefont{B.}~\bibnamefont{Munsky}},
  \bibinfo{author}{\bibfnamefont{I.}~\bibnamefont{Nemenman}}, \bibnamefont{and}
  \bibinfo{author}{\bibfnamefont{G.}~\bibnamefont{Bel}}, \bibinfo{journal}{The
  J. Chem. Phys} \textbf{\bibinfo{volume}{131}},
  (\bibinfo{year}{2009}).





\bibitem[{\citenamefont{Suter et~al.}(2011)\citenamefont{Suter, Molina,
  Gatfield, Schneider, Schibler, and Naef}}]{Suter2011Mammalian}
\bibinfo{author}{\bibfnamefont{D.~M.} \bibnamefont{Suter}},
  \bibinfo{author}{\bibfnamefont{N.}~\bibnamefont{Molina}},
  \bibinfo{author}{\bibfnamefont{D.}~\bibnamefont{Gatfield}},
  \bibinfo{author}{\bibfnamefont{K.}~\bibnamefont{Schneider}},
  \bibinfo{author}{\bibfnamefont{U.}~\bibnamefont{Schibler}}, \bibnamefont{and}
  \bibinfo{author}{\bibfnamefont{F.}~\bibnamefont{Naef}},
  \bibinfo{journal}{Science} \textbf{\bibinfo{volume}{332}},
  \bibinfo{pages}{472} (\bibinfo{year}{2011}).

\bibitem[{\citenamefont{Bialek}(2012)}]{Bialek2012Biophysics-Searching}
\bibinfo{author}{\bibfnamefont{W.}~\bibnamefont{Bialek}},
  \emph{\bibinfo{title}{Biophysics: Searching for Principles}}
  (\bibinfo{publisher}{Princeton University Press}, \bibinfo{year}{2012}).

\bibitem[{\citenamefont{Csan{\'a}dy et~al.}(2010)\citenamefont{Csan{\'a}dy,
  Vergani, and Gadsby}}]{Csanady2010Strict}
\bibinfo{author}{\bibfnamefont{L.}~\bibnamefont{Csan{\'a}dy}},
  \bibinfo{author}{\bibfnamefont{P.}~\bibnamefont{Vergani}}, \bibnamefont{and}
  \bibinfo{author}{\bibfnamefont{D.~C.} \bibnamefont{Gadsby}},
  \bibinfo{journal}{Proc. Natl. Acad. Sci. U.S.A.}
  \textbf{\bibinfo{volume}{107}}, \bibinfo{pages}{1241} (\bibinfo{year}{2010}).

\bibitem[{\citenamefont{Depken et~al.}(2013)\citenamefont{Depken, Parrondo, and
  Grill}}]{Depken2013Intermittent}
\bibinfo{author}{\bibfnamefont{M.}~\bibnamefont{Depken}},
  \bibinfo{author}{\bibfnamefont{J.~M.} \bibnamefont{Parrondo}},
  \bibnamefont{and} \bibinfo{author}{\bibfnamefont{S.~W.} \bibnamefont{Grill}},
  \bibinfo{journal}{Cell Reports} \textbf{\bibinfo{volume}{5}},
  \bibinfo{pages}{521 } (\bibinfo{year}{2013}).


\bibitem[{\citenamefont{Detwiler et~al.}(2000)\citenamefont{Detwiler,
  Ramanathan, Sengupta, and Shraiman}}]{Detwiler2000Engineering}
\bibinfo{author}{\bibfnamefont{P.~B.} \bibnamefont{Detwiler}},
  \bibinfo{author}{\bibfnamefont{S.}~\bibnamefont{Ramanathan}},
  \bibinfo{author}{\bibfnamefont{A.}~\bibnamefont{Sengupta}}, \bibnamefont{and}
  \bibinfo{author}{\bibfnamefont{B.~I.} \bibnamefont{Shraiman}},
  \bibinfo{journal}{Biophysical Journal} \textbf{\bibinfo{volume}{79}},
  \bibinfo{pages}{2801} (\bibinfo{year}{2000}).


\bibitem[{\citenamefont{Vaikuntanathan
  et~al.}(2013)\citenamefont{Vaikuntanathan, Gingrich, and
  Geissler}}]{Vaikuntanathan2013Dynamic}
\bibinfo{author}{\bibfnamefont{S.}~\bibnamefont{Vaikuntanathan}},
  \bibinfo{author}{\bibfnamefont{T.~R.} \bibnamefont{Gingrich}},
  \bibnamefont{and} \bibinfo{author}{\bibfnamefont{P.~L.}
  \bibnamefont{Geissler}},   \bibinfo{journal}{Phys. Rev. E} \textbf{\bibinfo{volume}{89}},
  \bibinfo{pages}{062108} (\bibinfo{year}{2014}).

\bibitem[{\citenamefont{Sartori et~al.}(2014)\citenamefont{Sartori, Granger,
  Lee, and Horowitz}}]{Sartori2014Thermodynamic}
\bibinfo{author}{\bibfnamefont{P.}~\bibnamefont{Sartori}},
  \bibinfo{author}{\bibfnamefont{L.}~\bibnamefont{Granger}},
  \bibinfo{author}{\bibfnamefont{C.~F.} \bibnamefont{Lee}}, \bibnamefont{and}
  \bibinfo{author}{\bibfnamefont{J.~M.} \bibnamefont{Horowitz}},
  \bibinfo{journal}{arXiv:1404.1027}  (\bibinfo{year}{2014}).

\bibitem[{\citenamefont{Barato et~al.}(2014)\citenamefont{Barato, Hartich, and
  Seifert}}]{Barato2014Efficiency}
\bibinfo{author}{\bibfnamefont{A.~C.} \bibnamefont{Barato}},
  \bibinfo{author}{\bibfnamefont{D.}~\bibnamefont{Hartich}}, \bibnamefont{and}
  \bibinfo{author}{\bibfnamefont{U.}~\bibnamefont{Seifert}},
  \bibinfo{journal}{arXiv:1405.7241}  (\bibinfo{year}{2014}).


\bibitem[{\citenamefont{Redner}(2001)}]{Redner2001A-Guide}
\bibinfo{author}{\bibfnamefont{S.}~\bibnamefont{Redner}},
  \emph{\bibinfo{title}{A Guide to First-Passage Processes}}
  (\bibinfo{publisher}{Cambridge}, \bibinfo{year}{2001}).


\bibitem[{\citenamefont{Blythe}(2001)}]{Blythe2001Nonequilibrium}
\bibinfo{author}{\bibfnamefont{R.~A.} \bibnamefont{Blythe}}, Master's thesis,
  \bibinfo{school}{Edinburgh} (\bibinfo{year}{2001}).

\bibitem[{\citenamefont{Hussain et~al.}(2014)\citenamefont{Hussain, Gupta,
  Hirning, Ott, Matthews, Josi{\'c}, and Bennett}}]{Hussain2014Engineered}
\bibinfo{author}{\bibfnamefont{F.}~\bibnamefont{Hussain}},
  \bibinfo{author}{\bibfnamefont{C.}~\bibnamefont{Gupta}},
  \bibinfo{author}{\bibfnamefont{A.~J.} \bibnamefont{Hirning}},
  \bibinfo{author}{\bibfnamefont{W.}~\bibnamefont{Ott}},
  \bibinfo{author}{\bibfnamefont{K.~S.} \bibnamefont{Matthews}},
  \bibinfo{author}{\bibfnamefont{K.}~\bibnamefont{Josi{\'c}}},
  \bibnamefont{and} \bibinfo{author}{\bibfnamefont{M.~R.}
  \bibnamefont{Bennett}}, \bibinfo{journal}{Proceedings of the National Academy
  of Sciences} \textbf{\bibinfo{volume}{111}}, \bibinfo{pages}{972}
  (\bibinfo{year}{2014}).

\bibitem[{\citenamefont{Segel}(1975)}]{Segel1975Enzyme}
\bibinfo{author}{\bibfnamefont{I.~H.} \bibnamefont{Segel}},
  \emph{\bibinfo{title}{Enzyme Kinetics}}, vol. \bibinfo{volume}{360}
  (\bibinfo{publisher}{Wiley, New York}, \bibinfo{year}{1975}).

\end{thebibliography}
\end{document}